\documentclass[twocolumn]{aastex631}
\journalinfo{}
\shorttitle{Growth in Spite of Recycling}
\shortauthors{Bailey \& Zhu}

\begin{document}
\title{Growing Planet Envelopes in Spite of Recycling Flows}

\correspondingauthor{Avery Bailey}
\email{avery.bailey@unlv.edu}

\author[0000-0002-6940-3161]{Avery P. Bailey}
\affiliation{Department of Physics and Astronomy, University of Nevada, Las Vegas, 4505 South Maryland Parkway, Las Vegas, NV 89154-4002, USA}
\affiliation{Nevada Center for Astrophysics (NCfA), University of Nevada, Las Vegas, NV, USA}

\author[0000-0003-3616-6822]{Zhaohuan Zhu}
\affiliation{Department of Physics and Astronomy, University of Nevada,
Las Vegas, 4505 South Maryland Parkway, Las Vegas, NV 89154-4002, USA}
\affiliation{Nevada Center for Astrophysics (NCfA), University of Nevada, Las Vegas, NV, USA}



\begin{abstract}
The hydrodynamic exchange of a protoplanet’s envelope material with the background protoplanetary disk has been proposed as one mechanism to account for the diversity of observed planet envelopes which range in mass fractions of $\sim 1\%$ for super-Earths to $\sim 90\%$ for giants. Here we present and analyze 3D radiation-hydrodynamics models of protoplanet envelopes to understand how the exchange of mass and energy between the protoplanet and background disk influences the formation process. Our protoplanet envelope simulations show an exchange of material bringing the outer $\gtrsim 0.4R_b$ envelope to steady state. This exchange provides a continuous source of energy, which acts to increase the observed luminosity beyond that inferred from the binding energy liberated from Kelvin-Helmholtz contraction alone -- a finding important for potential protoplanet observations. The inner $\lesssim 0.4R_b$, on the other hand, appears insulated -- growing in accordance with 1D quasi-static theory. We incorporate these 3D hydrodynamic effects into an extensible 1D framework with a physically motivated three-layer recycling parameterization. Specializing to the case of Jupiter, recycling produces minimal changes to the growth rate with the planet still entering runaway accretion and becoming a gas giant in $\sim 1$ Myr. Even in the inner disk ($0.1$ AU), our 1D models suggest that recycling is not so robust and ubiquitous as to stop all cores from becoming giants. At the same time however, this recycling can delay a runaway phase by an order-of-magnitude depending on the inner disk conditions and core mass.
\end{abstract}



\section{Introduction} \label{sec:intro}
The core accretion theory of planet formation exists as a favorable mechanism to form gas giants akin to Jupiter.
In the core accretion model, gas accretion and the consequent setting of 
planet formation timescale is mediated by the Kelvin-Helmholtz contraction of the protoplanet envelope \citep{Mizuno1980, Pollack+1996}. 
Traditionally, this slow contraction has been modeled as a 1D quasi-static process where the envelope structure is 
considered to be nearly hydrostatic with small deviations arising from gravitational contraction which 
drive the long term envelope evolution \citep{BodenheimerPollack1986, Ikoma+2000}. 
Historically core accretion and its subsequent extensions (e.g. pebble accretion) 
have been the de-facto standard for solar system formation and planet formation in general. To its credit, core accretion 
naturally explains the metal-rich interiors of Jupiter and Saturn \citep{Guillot+1997, Wahl+2017, Militzer+2022} 
and the correlation between giant planet occurrence and host star metallicity \citep{FischerValenti2005}.
More recently, observational challenges 
and theoretical considerations (driven by increased computing power and advancement in the methods for numerical solution of the 
fluid equations) have prompted a re-examination of the quasi-static core accretion framework.

From an observational point of view there appear at least three pieces of evidence incompatible with the traditional picture of 
core accretion. First is the existence of planets with core masses far exceeding what is expected from the core accretion 
scenario. HD 149026b, for example, is a transiting Saturn mass planet with unusually high density that internal models would estimate as 
having a core of $\sim 70 M_\odot$. While the existence of such a core argues for a core accretion origin over something like 
gravitational instability, the inferred core mass far exceeds the relatively robust 5-20 $M_\oplus$ critical core mass predicted by 
quasi-static core accretion models \citep{PisoYoudin2014, Piso+2015}. Second is the existence of numerous super-Earths and 
mini-Neptunes in exoplanet systems. These planets have atmosphere to core mass ratios at the order of $1\%$, massive enough 
to suggest they formed in the gaseous protoplanetary disk. At they same time, they obviously did not runaway to become gas giants 
even though many have core masses larger than the theoretical critical core mass. Finally is an observed abundance of 
sub-Saturn mass planets compared to the predictions of core accretion. In theory, there ought to be a bimodal distribution 
of planets consisting of a population of gas giants that were massive enough to enter runaway accretion and 
a low-mass population that failed to runaway before the disk dispersed with the minimum occurrence rate landing 
in the sub-Saturn/Neptune regime. While population synthesis models exhibit some indication of this bimodality 
\citep{IdaLin2004, Mordasini2018}, sub-Saturns appear to be nearly as common as their 
Jovian counterparts \citep{DongZhu2013,Petigura2018}. 

Several solutions have been put forth to address these observational tensions. Late formation or high opacity would 
allow super-Earths to avoid the transition to runaway accretion \citep{Lee+2014, LeeChiang2015}. A hierarchical merging 
formation channel where cores grow through giant impacts \citep{IkomaHori2012, Ginzburg+2016} or gap-opening \citep{FungLee2018, 
GinzburgChiang2019} would similarly slow gas accretion and ease these tensions. 
The advective transport of energy and mass between the protoplanet envelope and the protoplanetary disk, 
termed atmospheric recycling, has also been advanced as a mechanism to explain the abundance of sub-Saturns \citep{Lee2019} 
and super-Earths \citep{Ormel+2015, Cimerman+2017, Ali-Dib+2020, Moldenhauer+2021, Moldenhauer+2022} and is the primary 
subject of this paper.

From a theoretical perspective, some recycling must occur as all protoplanet envelopes are embedded in the Keplerian 
shear of a background protoplanetary disk at some point in their evolution. As an inherently multi-dimensional phenomenon 
however, only recently, with advances in computing power and techniques has it been possible to simulate this 
recycling interaction in 3D with the appropriate physics. While much impressive work has been done on the subject, 
the full scope of recycling and its application to the full diversity of planet forming environments is not entirely understood. 
If super-Earths and gas giants are thought to form in the same core accretion framework with recycling stalling the 
super-Earth evolution but not the giants for example, there must be some differential recycling efficiency or preferential 
behavior surrounding the circumstances of formation. 
The magnitude of this efficiency can be best understood from 3D radiation-hydrodynamical simulations which themselves 
suffer from short simulation times ($\sim 10-10^3$ orbits) compared to the evolutionary timescale. This necessitates 
careful methods and analysis to properly interpret 3D models in the context of long-term evolution and eventual formation 
outcome. In this work we examine and refine some of our existing 
3D radiation-hydrodynamics models of protoplanetary envelopes, taking care to study the mass and energy budget of the 
envelope as these diagnostics inform the long term evolution of these protoplanets.

In section \ref{sec:methods} we present numerical methods and an overview of the dynamics present in our  
radiation-hydrodynamics envelope models. Section \ref{sec:mass} quantifies the mass exchange
between planet and background disk by examining the envelope mass budget and tracking the 
circumplanetary flow topology with passive scalars. Because hydrodynamic flows also transfer energy 
between envelope and disk, we also examine the envelope energetics and present them in section \ref{sec:energy}.
In section \ref{sec:disc} we summarize our findings and interpret these models in the 
context of core accretion and the formation of planets like Jupiter along with close-in super-Earths/mini-Neptunes.

\section{Simulation Details}\label{sec:methods}
The simulations and setup employed here are similar to those of \citet{Bailey+2023} but 
generally at higher resolution, extended simulation time, and with the inclusion of passive scalars. 
We summarize the most important information here but additional details may be found in 
the preceding work. The radiation-hydrodynamics simulations are performed with \textsc{Athena++} 
in a local frame centered on the planet with boundary conditions fixed to those of the background disk. 
The mass and momentum equations for this system are the standard inviscid hydrodynamics ones 
with rotational source terms arising from Coriolis and centrifugal forces along with gravitational source terms 
from the planet and central star. The result is a stratified shearing-box with a planet potential (though 
with fixed boundaries rather than shear-periodic). Radiative transfer is calculated with a formal solution of 
the transfer equation:
\begin{equation}
\frac{dI}{d\tau} = S-I
\end{equation}
with a short characteristics method \citep{Davis+2012}. Radiation is then coupled to the hydrodynamics by 
computing $J=\int Id\Omega$ and adding as a source term in the dimensionless energy equation:
\begin{equation}
\frac{\partial E}{\partial t} + \nabla\cdot\left(Ev + pv\right) = -\rho v\cdot\nabla\Phi + \rho\kappa\beta(J-S).\label{eq:energy}
\end{equation}
We use gray, constant opacities with a Planck source function $S = (p/\rho)^4$ for the radiative 
transfer in these models. All simulations are run with static mesh refinement and a maximum resolution of 
$256$ cells$/H_0$ until at least time $t=400\Omega_0^{-1}$. 

\subsection{Model Parameters}
The simulations adopt dimensionless units where disk scale height, orbital frequency, and midplane 
gas density are set to unity ($H_0=\Omega_0=\rho_0 = 1$). A given model is then determined by the four 
dimensionless parameters: $q_t, \kappa, \beta, \epsilon/H_0$ which are explained in more physical terms below.
We use the subscript $0$ to refer to characteristic dimensional values of the system which, by virtue of setting
$H_0=\Omega_0=\rho_0 =1 $, can be used to dimensionalize the dimensionless variables presented for the remainder 
of the this work.

\begin{itemize}
\item The planet core's thermal mass $q_t$:
\begin{equation}\label{eq:dim1}
q_t = \frac{M_c}{M_\ast}\left(\frac{H_0}{a_0}\right)^{-3}.
\end{equation}
This characterizes the core-to-star mass ratio in a way that quantifies the strength of perturbation 
the planet exerts on the background protoplanetary disk. Subthermal planets ($q_t \lesssim 1$)  
weakly perturb the disk whereas superthemal planets ($q_t \gtrsim 1$) have masses and Hill radii 
large enough to strongly perturb the surrounding disk.
\item The dimensionless opacity $\kappa$
\begin{equation}
\kappa = \kappa_0 \rho_0 H_0.
\end{equation}
Because simulations are initialized as an isothermal stratified disk, the dimensionless $\kappa$ can also be 
considered as the optical depth to the midplane of the background disk: $ \kappa\sim 0.8\tau_{\rm midplane}$
\item The dimensionless gas-radiation coupling $\beta$ 
\begin{equation}\label{eq:dim3}
\beta = \frac{ac}{c_0}\left(\frac{\mu m_p}{k}\right)\frac{T_0^3}{\rho_0}.
\end{equation}
This parameter encodes the thermal properties of the local disk and, along with $\kappa$, couples the radiative 
transfer to the gas energy. Note that $\beta$ can also be written 
as the ratio of energy fluxes characteristic to the radiative and gas properties of the disk
\begin{equation}
\beta = \frac{c}{c_0}\frac{a T_0^4}{p_0}\sim\frac{F_{\rm rad}}{F_{\rm gas}},
\end{equation}
and can also be found in \citet{Jiang+2012} as $\mathbb{CP}$ (see their Eq. 1).
\item The gravitational softening length $\epsilon/H_0$ defined by the adopted planet potential:
\begin{equation}
\Phi_p = -\frac{GM_c}{\sqrt{r_0^2+\epsilon^2}}.
\end{equation}
This is a parameter employed for numerical considerations. Ideally it would be set to match the core 
radius of a forming planet. For planets like a proto-Jupiter however, this scale lies orders of magnitude below 
the grid scale and would be computational infeasible. Therefore we artificially set the parameter to scale with 
the gravitational potential $\epsilon/H_0 = q_t/10$ and study the outer envelopes of such systems. While less physical,
this choice at least ensures that the models are fully resolved numerically.
\end{itemize}

For the models in this work we restrict the parameter space to models with $\beta=1$ and test two different opacities with 
$\kappa=(1,100)$, and three masses with $q_t=(0.5,1,2)$. We also isolate the role of background shear and diagnose systematics 
of these models by running some simulations designated \texttt{-noshear}. These models have rotational source terms omitted 
from the momentum equation and a zero-velocity initial condition to simulate the case of a core immersed in a static stratified medium.
All models presented in this work and their corresponding dimensionless parameters are summarized in Table \ref{tab:models}.
Also provided in the table are possible choices of physical planet parameters at $\sim 0.1$ and $\sim 5$au (assuming a solar-mass host star) that coincide with the chosen dimensionless parameters.
It is important to recognize that the displayed physical parameters are not the only acceptable choices and that 
quantities can be scaled up or down so long as they satisfy equations \ref{eq:dim1}-\ref{eq:dim3} for a given set of 
dimensionless parameters. 

\begin{deluxetable*}{lcccc|cccc|cccc}
\tablewidth{0pt} 
\tablecaption{Table of Model Parameters}
\tablehead{
\colhead{} & \multicolumn{4}{c}{Dimensionless Values} & \multicolumn{4}{c}{Consistent $\sim 5$au  Values} & \multicolumn{4}{c}{Consistent $\sim 0.1$au Values}\\
\colhead{Name} & \colhead{$q_t$} & \colhead{$\beta$} & \colhead{$\kappa$} & \colhead{$\epsilon/H_0$} & 
\colhead{$\Sigma_0$} & \colhead{$T_0$}  & \colhead{$\kappa_0$} & \colhead{$M_c$} &
\colhead{$\Sigma_0$} & \colhead{$T_0$} & \colhead{$\kappa_0$} & \colhead{$M_c$} \\
\colhead{} & \colhead{} & \colhead{} & \colhead{} & \colhead{} &
\colhead{(g/cm$^2)$} & \colhead{(K)} & \colhead{(cm$^2$/g)} & \colhead{($M_\oplus$)} &
\colhead{(g/cm$^2$)} & \colhead{(K)}  & \colhead{(cm$^2$/g)} & \colhead{($M_\oplus$)}
} 
\startdata 
\texttt{b1k1}                   & 0.5 & 1 & 1 & 0.05 & 250 & 65  & 0.01 & 8 & $1.5\times 10^4$ & 1800 & $0.017$ & 3 \\
\texttt{b1k1-noshear}     & - & - & - & - & - & - &  - & -   & - & - & - & -   \\
\texttt{b1k1-q1}              & 1 & - & - & 0.1 & - & - & - & 16  & - & -  & - & 6  \\
\texttt{b1k1-q2}              & 2 & - & - & 0.2 & - & - & - & 32  & - & -  & - & 12  \\
\texttt{b1k100}               & 0.5 & - & 100 & 0.05 & - & - & 1 & 8  & -  & - & 1.7 & 3   \\
\texttt{b1k100-noshear} & - & - & - & - & - & - & - & - & - & - & - & -  
\enddata
\tablecomments{cells marked with - are assumed to have the value of the preceding row}
\label{tab:models}
\end{deluxetable*}

\subsection{Flow Structure}
\begin{figure*}
\plotone{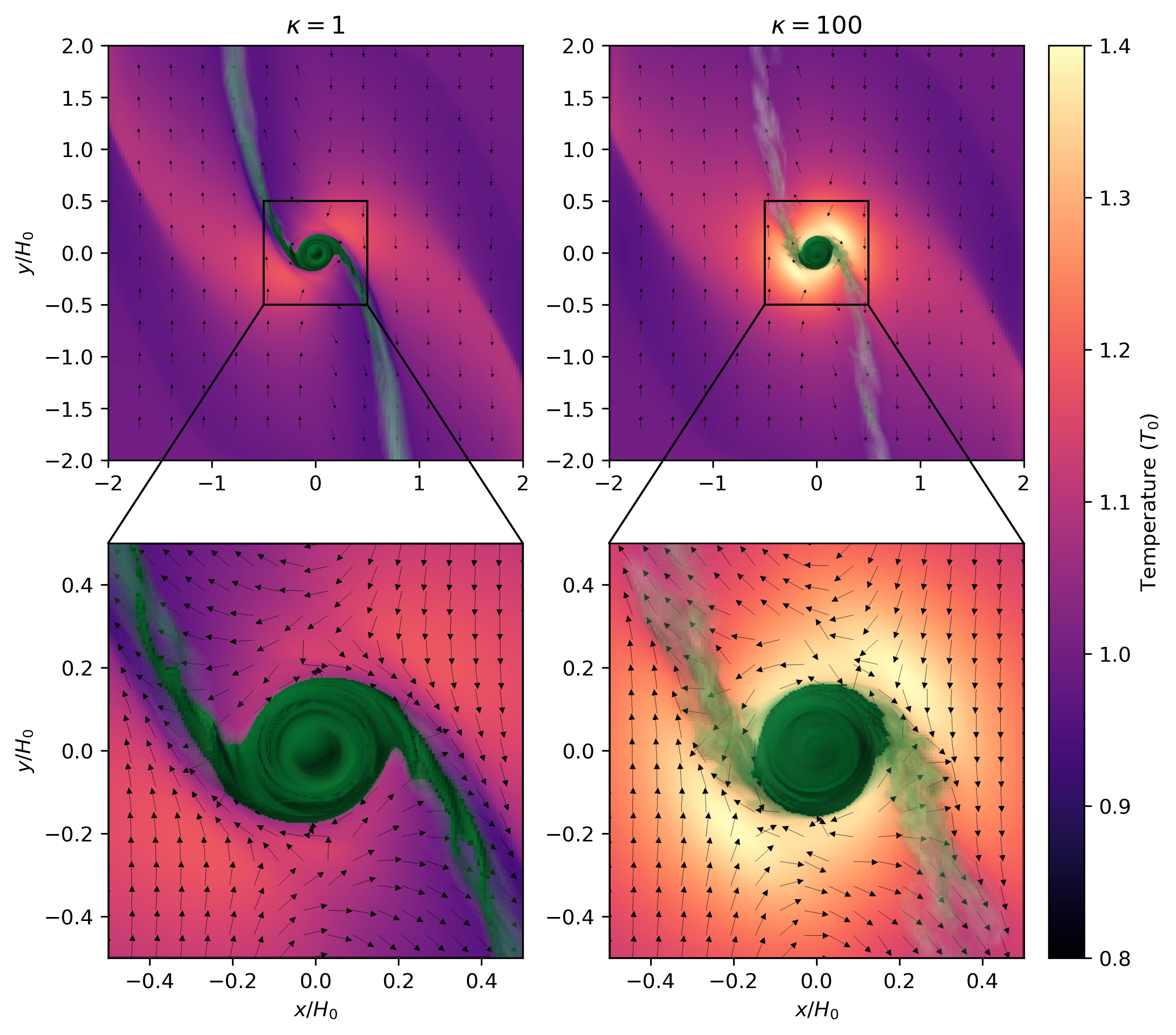}
\caption{Top-down snapshots of simulations \texttt{b1k1} and \texttt{b1k100} taken at $t = 400\Omega_0^{-1}$. 
In green is a volume rendering of passive scalars but on a greatly exaggerated scale. In reality, the spiral streamers 
are orders of magnitude lower density than the core material. The scalars are superimposed upon a slice of the 
simulation midplane where temperature and velocity directions are displayed.\label{fig:midplane}}
\end{figure*}

To provide an example of the setup and value of these 
models, two snapshots taken at the end of the fiducial simulations \texttt{b1k1} and \texttt{b1k100} are shown in Fig.\@ {\ref{fig:midplane}}.
The top panels display the full simulation width while the bottom panel zoom-in 
to the scale of the planet's Bondi radius $R_b$. Passive scalars shown in green allow for a visualization of 
the recycling of circumplanetary material along horseshoe orbits back to the protoplanetary disk. While the 
optical properties of the volume rendered scalars are the same between both displayed models, the scaling is 
exaggerated relative to scalar density to highlight the recycling process. In fairness, the tenuous streams of 
recycled material are substantially lower density than the saturated inner core. Also shown is the midplane 
temperature structure and flow field which appears similar to the results obtained in other works. 

These simulations, even with more sophisticated thermodynamics, retain a qualitatively similar flow field to other $\beta$ cooling models \citep{KurokawaTanigawa2018} or even some isothermal ones \citep{Ormel+2015, Fung+2015, Bailey+2021}.
In the midplane, at larger lateral distance the flow reduces to retrograde oriented shearing streamlines of the background disk. 
Along $x=0$, the models recover the typical family of horseshoe streamlines. 
Closer to the planet are recycling flows sourced from horseshoes at higher altitude.
These outflows through which material is being returned to the background disk are markedly cooler than the background equilibrium and lie just interior to the heated spiral arms.
Though this cooling is far more obvious in the $\kappa=1$ case.

\begin{figure*}
\plotone{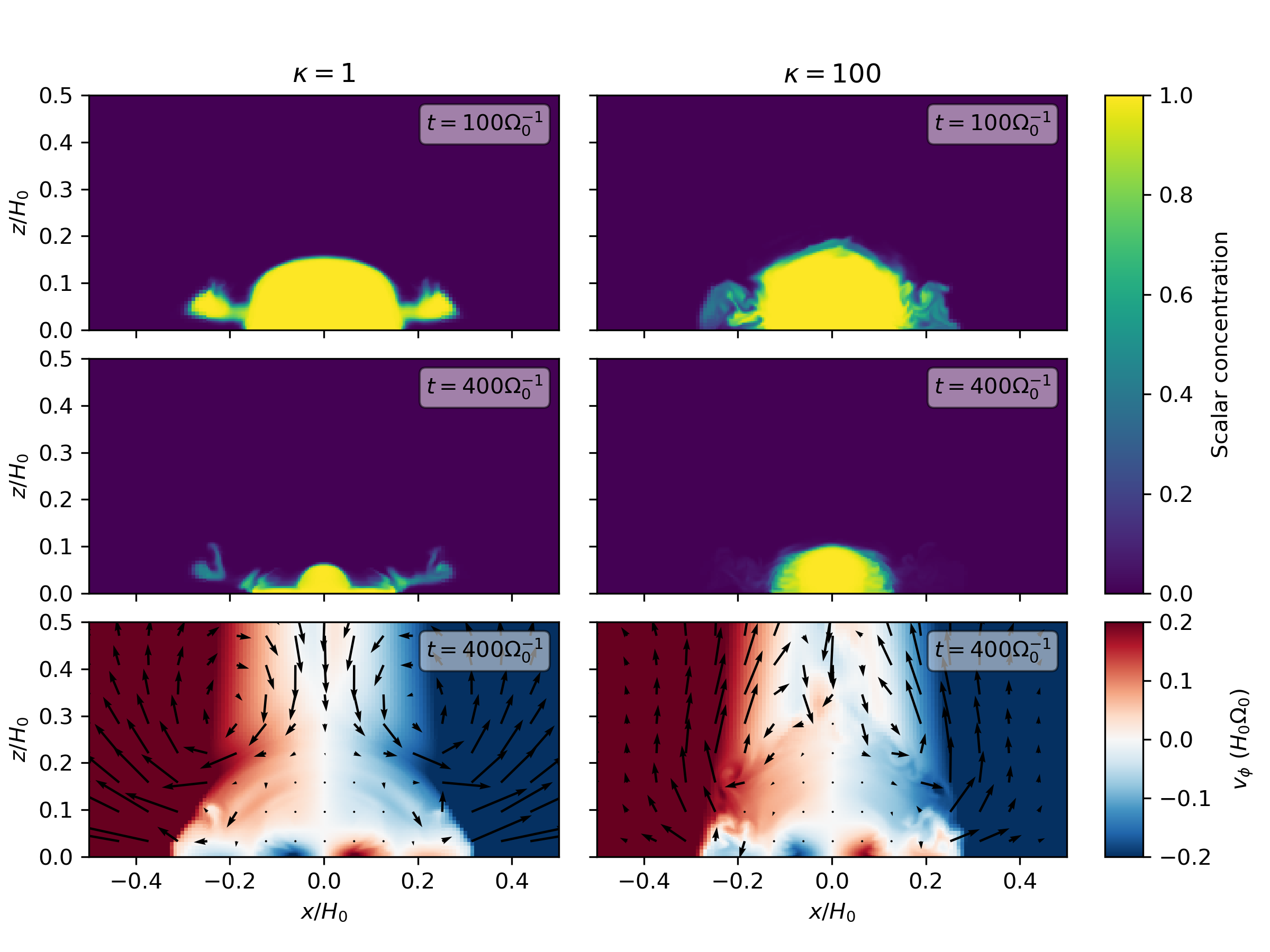}
\caption{Vertical slices showing the evolution of scalar concentrations for models \texttt{b1k1} and \texttt{b1k100}. 
Corresponding toroidal and poloidal (vectors) velocities at $t=400\Omega_0^{-1}$ are plotted in the bottom panel.
 \label{fig:vertscalar}}
\end{figure*}

The circumplanetary flow patterns 
at least qualitatively also match those found in other radiation-hydrodynamics models of atmospheric recycling
\citep{Moldenhauer+2021, Moldenhauer+2022}. Akin to other works, we fill the planet's Bondi sphere with passive scalars to 
visualize the recycling. Because our simulations are not in thermodynamic equilibrium however, the passive scalars are 
also tracking the accretion process, lessening their overall effectiveness as a recycling diagnostic. Upon injecting 
the passive scalars ten orbits after simulation start, material outside of $R_b/2$ in all simulations is recycled within 
several orbits due to the short recycling timescales there. After $400\Omega_0^{-1}$, as is shown in Figures \ref{fig:midplane}
\& \ref{fig:vertscalar}, passive scalars remain only within the innermost $R_b/5$.
We emphasize however that because the models are not in thermodynamic equilibrium $R_b/5$ is not representative of a characteristic recycling radius. 
While it may appear that passive scalars in the region between $R_b/5 < r< R_b/2$ have disappeared between $t=100\Omega_0^{-1}$ and $400\Omega_0^{-1}$ perhaps due to recycling, there is also ongoing accretion. 
Rather than being lost to recycling, said scalars have been accreted radially into the core.
Indeed, out of the scalar mass within $R_b/5$ at simulation end, roughly one-fourth to one-third is additionally accreted scalar material with the rest being unrecycled 
mass remaining from the initial injection. We also find that while scalar material is predominantly lost to the background disk 
via the $x$ and $y$-directions, accreted scalar fluxes are largest in the vertical direction -- reminiscent of the pattern of meridional 
circulation. Between the different opacity models, the passive scalar field appears qualitatively similar even though the optically 
thicker case exhibits small scale turbulent velocities. The optically thinner case does show stronger retention of passive scalars near the midplane, 
a feature also appearing in the \citet{Moldenhauer+2022} models. 

Apart from the midplane, the models exhibit flow aspects significantly different from existing cases however -- aspects which we believe ultimately impact the implications for planet formation.
Viewing the vertical structure of the velocity field in Fig.\@ \ref{fig:vertscalar}, we see similar toroidal rotation patterns independent of the opacity. 
The circumplanetary region, sourced from high altitude prograde oriented streamlines, retains a prograde rotating atmosphere that is highly sub-Keplerian. 
The poloidal velocities however transition from strong polar inflow and are dramatically damped within the inner $R_b/2$. 
This appears reminiscent of the work of \citet{KurokawaTanigawa2018} where envelope models that include cooling, establish a strong entropy gradient in the atmosphere. 
This entropy gradient results in an additional buoyant force upon a fluid element impinging upon the inner envelope and creates a barrier to recycling.
This in in contrast to both adiabatic and isothermal models which do not have cooling and thus do not set up strong entropy gradients.
In these models, recycling appears much more capable in accessing the inner envelope which would act to interfere with the quasi-static contraction process.
Though we use a more realistic cooling by doing the radiative transfer, the buoyancy barrier phenomenon put forth by \citet{KurokawaTanigawa2018} appears consistent with the models presented here.

\section{Mass Exchange}\label{sec:mass}
Recycling flows tend to exchange high entropy disk material with the cooled low entropy 
gas within the planet atmosphere. This effectively heats the outer regions of the planetary envelope to 
an equlibrium where mass can no longer be accreted. Within this context, it is natural to examine the properties of
mass accretion in our models and study locations where accretion has been inhibited to determine the 
influence recycling flows have on envelope growth. 

\begin{figure*}
\plotone{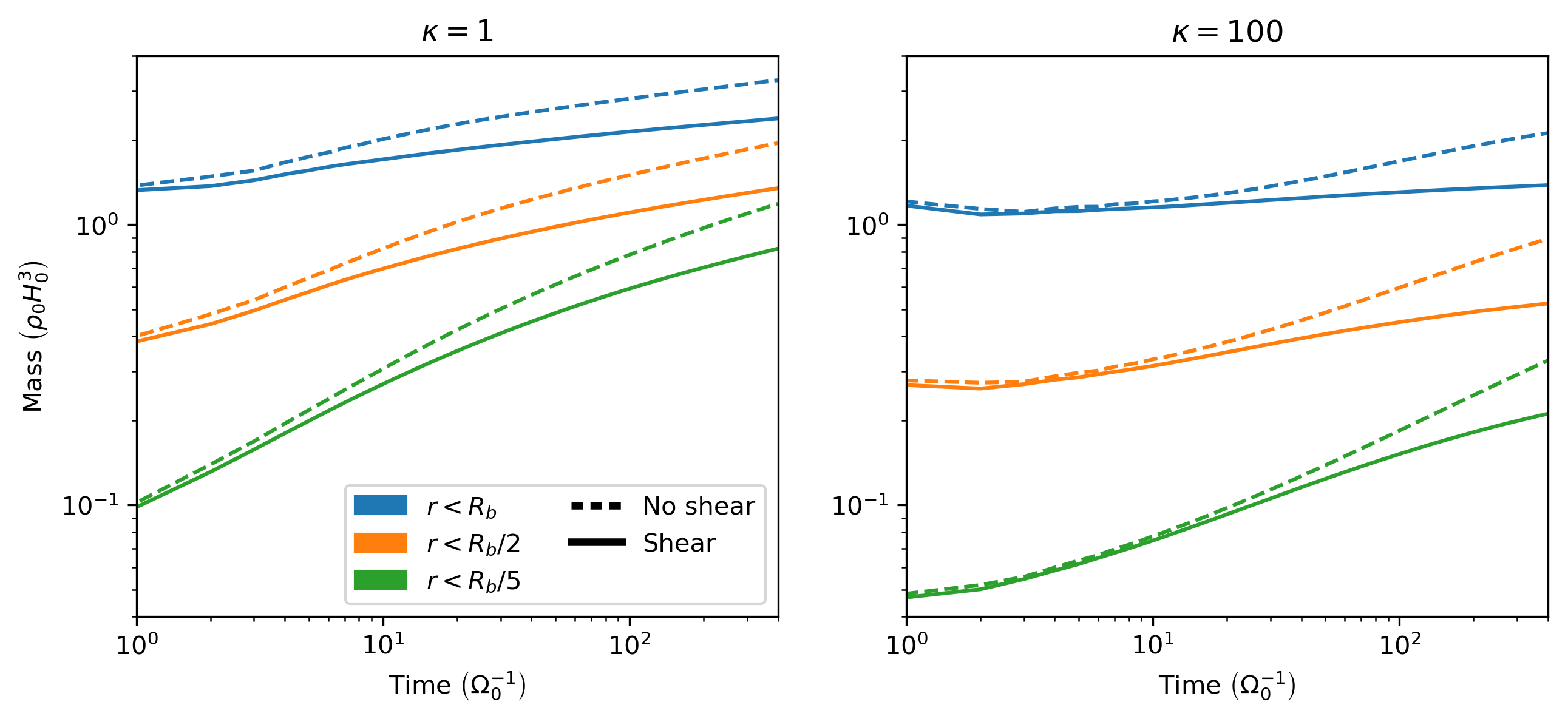}
\caption{Evolution of the envelope mass as a function of time for models \texttt{b1k1} and \texttt{b1k100}. 
Three choices of envelope radius ($r = R_b$, $R_b/2$, $R_b/5$) are shown and compared against the mass in an 
equivalent \texttt{-noshear} model.\label{fig:mcurves}}
\end{figure*}

In Figure \ref{fig:mcurves} we plot the mass interior to several 
radii for the fiducial \texttt{b1k1} and \texttt{b1k100} models. 
At early times, on the dynamical timescale, the mass 
increase occurs as the envelope reaches an approximate hydrostatic equilibrium in response to the newly 
introduced planet potential. On longer timescales, radiative cooling in the interior drives the additional accretion of mass. 
Notably, we see that the envelope masses measured are still increasing even at the end of the fiducial 
runtime $t=400\Omega_0^{-1}$. This contrasts the simulations of \citet{Moldenhauer+2022} where thermodynamic equilibrium 
and steady state are achieved after $\sim 300\Omega_0^{-1}$.
We also emphasize that the relative mass change in logspace at the largest scale (blue curves) is small but the relative mass change at the small scale (green curves) is large, suggesting that significant mass accretion is occurring in the interior envelope.

\begin{figure*}
\plotone{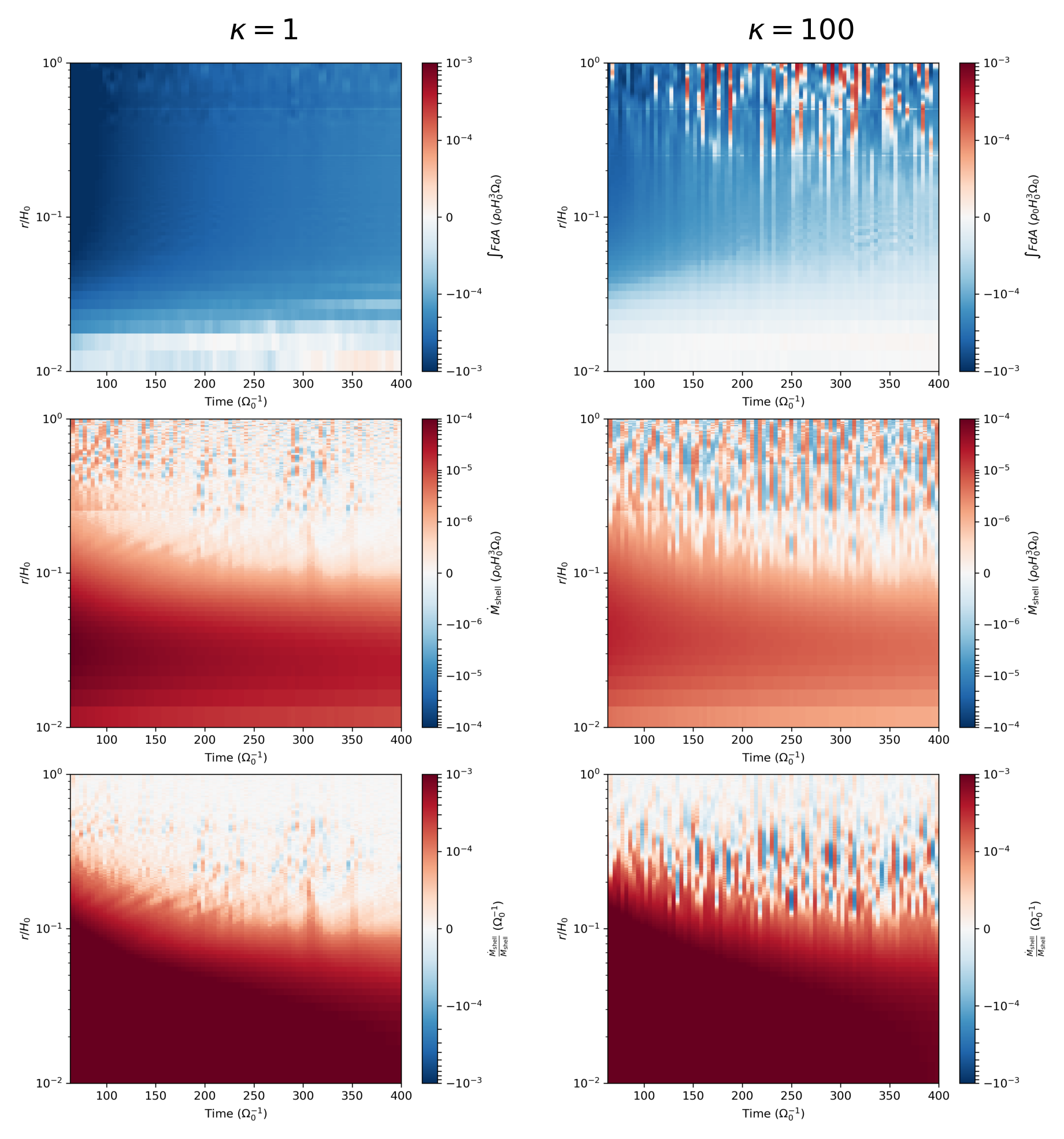}
\caption{Top: integrated mass fluxes through a surface located at $r$ for models \texttt{b1k1} and \texttt{b1k100}. 
Middle: $\dot{M}$ of a shell of single grid cell thickness located at $r$. 
Bottom: The corresponding $\dot{M}_{\rm shell}/M_{\rm shell}$ at $r$.
The geometry in all cases is a box of 
half-width $r$ to match the cartesian simulation domain.\label{fig:mflux}}
\end{figure*}

While Fig.\@ \ref{fig:mcurves} demonstrates that envelope masses are increasing for the duration of the simulations, it is not 
sufficient to determine the extent of the envelopes. After all, the mass increases for all curves could be a reflection of only the 
innermost regions accreting. To address this, we present Fig.\@ \ref{fig:mflux} in which integrated net mass fluxes at a given distance 
from the center of the simulation domain (i.e.\@ planet center) are mapped out as a function of time. 
Moving outwards from the planet center we construct concentric boxes aligned on cell interfaces and calculate the net mass fluxes through 
each box. These integrated fluxes are plotted in the upper panels of Fig.\@ \ref{fig:mflux} with the ordinate giving 
the half-width of the box (in units of $H_0$) used to perform the surface integral. Locations marked blue possess a net inward flux 
of mass at corresponding time while those marked red show an outward flux. And by the divergence theorem, 
these fluxes are equal albeit opposite to the net accretion rate $\dot{M}$ taking place within the constructed box. 

We find that the fluxes generally end up pointing inward everywhere, a finding reflected by the positive derivatives of the 
mass curves in Fig.\@ \ref{fig:mcurves}. The full flux map of Fig.\@ \ref{fig:mflux} however, also contains gradients in 
the $r/H_0$ direction, suggesting variations in the mass accretion rate for different distances from the planet core.
For the $\kappa=1$ model, the fluxes appear nearly constant at large distances from the planet, suggesting no net 
mass accretion in those regions. For more interior regions where gradients in the flux appear steeper there is greater 
associated mass accretion. The $\kappa=100$ model appears to show more extended gradients in flux, but as we show 
in further analysis, these are more an artifact of the \texttt{b1k100} fluxes existing in the linear part of the log-linear 
color normalization than a true difference between the models. 
Similarly the near zero flux at $r\sim 0.01$ appears to be an artifact from our post-processing of the fluxes that arises when close to the grid scale ($\sim 1/256$).

Because density changes are related to fluxes through the continuity equation $\dot{\rho} = -\nabla\cdot F$, 
the local accretion rates can be better presented and quantified if instead of considering the fluxes, flux gradients are considered. 
This is reflected in the middle panels of Fig.\@ \ref{fig:mflux} where plotted are the differences in fluxes between the adjacent 
cell interfaces plotted in the upper panels. By the continuity equation, these plotted flux differences are equivalent to the mass accretion 
rates $\dot{M}_{\rm shell}$ in concentric shells of single cell width thickness. 
To make the measurement independent of resolution, we also supply $\dot{M}_{\rm shell}/M_{\rm shell}$ in the lower panels of Fig.\@ \ref{fig:mflux}.
Physically this would represent some inverse mass accretion timescale for a given radial location.
The middle and lower panels of Fig.\@ \ref{fig:mflux} show that 
by the end of simulation runtime, there is an inner core where $\dot{M}$ is solidly red and still increasing. Exterior, there is a recycled region 
consistent with zero accretion. This outer recycled region is not quiescent but shows temporal variation between local 
net mass accretion and loss. The variation appears to be due to unsteady fluid motion in the envelope also reflected in 
the turbulent looking outflow streams in Fig.\@ \ref{fig:midplane}.

\begin{figure}
\plotone{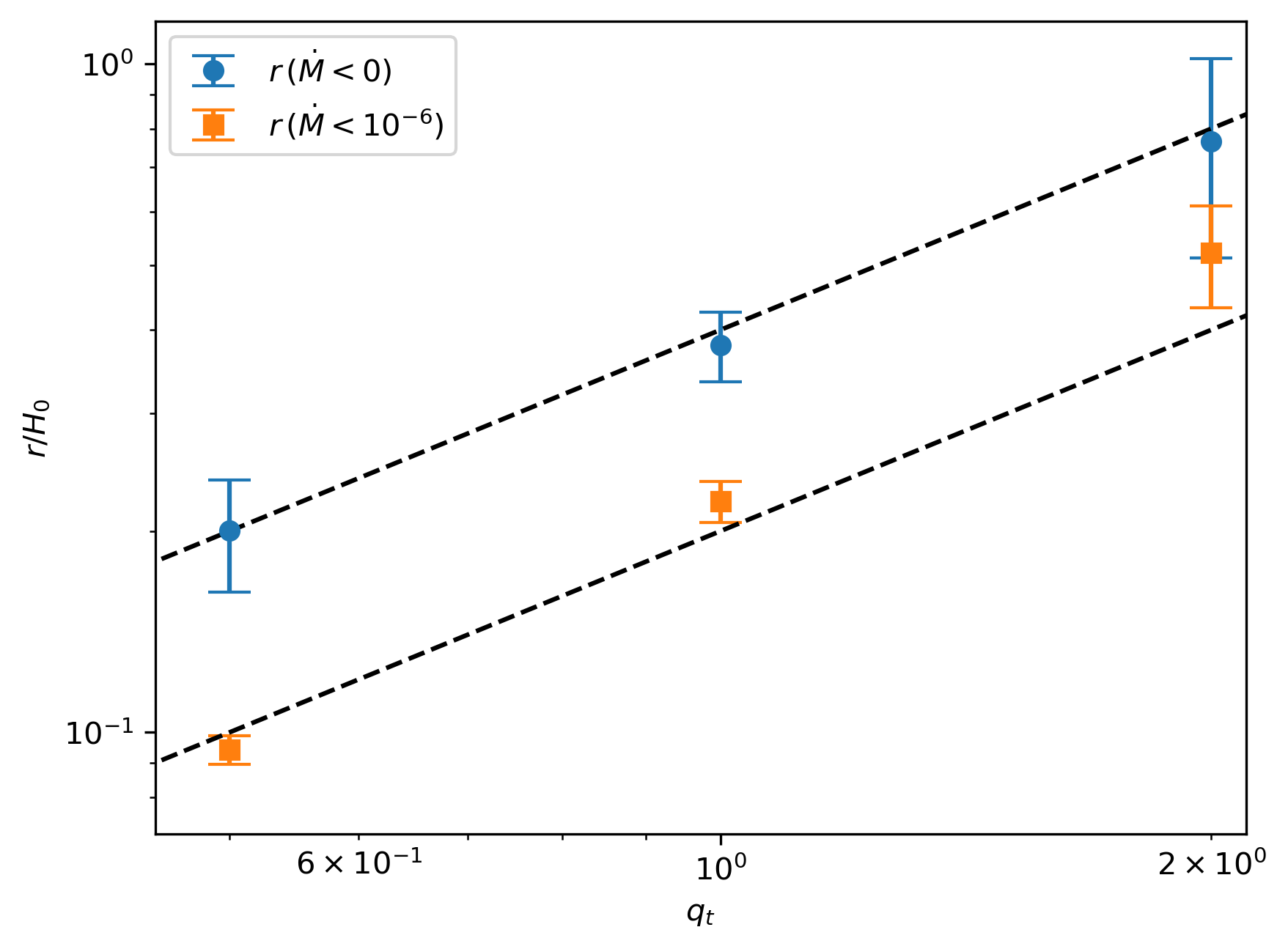}
\caption{Box half-widths where $\dot{M}_{\rm shell}$ vanishes (blue) or at least becomes less than $10^{-6}$ (orange). 
Values for $r/H_0$ come from models \texttt{b1k1}, \texttt{b1k1-q1}, \texttt{b1k1-q2} averaged over the last $100\Omega_0^{-1}$ 
with errorbars giving the standard deviation over the final $100\Omega_0^{-1}$ and therefore some estimate of the time variability.
Black dashed lines are drawn as references for linear scaling $r/H_0\propto q_t$\label{fig:q}}
\end{figure}

The main advantage of Fig.\@ \ref{fig:mflux} is that the transition region from accreting to non-accreting can be 
quantified. To define this transition radius for a given simulation snapshot we take the minimum radius at which a shell 
switches to mass loss, effectively taking a contour on Fig.\@ \ref{fig:mflux} for $\dot{M}_{\rm shell}=0$. With this
definition, we find that by the end of simulation the transition radius levels out to a value $r_t \sim H_0/5 \sim 0.4 R_b$ for both 
simulations depicted in Fig.\@ \ref{fig:mflux}. Repeating this procedure for our other models, we find that the transition 
radius scales linearly with $q_t$ and therefore linearly with the planet mass. These transition radii are plotted accordingly in 
Fig.\@ \ref{fig:q}. 
To give some account of the variability, we average the value of transition radius over the final $100\Omega_0^{-1}$ 
and present the standard deviation as error bars. In addition to points for $\dot{M}_{\rm shell}=0$, points for the radius where 
$\dot{M}_{\rm shell} = 10^{-6}$ are also plotted in Fig.\@ \ref{fig:q}. While less physically relevant, the $\dot{M}_{\rm shell}=10^{-6}$ 
boundary is where the color-scale in Fig.\@ \ref{fig:mflux} switches from log to linear and therefore can be readily discerned by eye 
in Fig.\@ \ref{fig:mflux}. The $\dot{M}_{\rm shell}=10^{-6}$ transition location shows a similarly approximate linear scaling with 
the thermal mass demonstrating the insensitivity of the results to a precise definition of this transition radius. 



\section{Energetics}\label{sec:energy}
In addition to transferring mass, atmospheric recycling can transfer energy between the forming envelope 
and background disk. Because accretion is governed by the cooling timescale for the bulk of protoplanet 
lifetimes, one can also hope to study the envelope accretion process by looking at the envelope energetics. 
In this section we study the energy budget of our simulations by considering three energy sources/sinks and 
testing systematics against our \texttt{-noshear} models. 
The energy sources considered come from a rewrite 
of the energy conservation equation (\ref{eq:energy}). 
To do so, one can split the potential into planetary and stellar components $\Phi = \Phi_p + \Phi_s$ where 
$\Phi_p = q_t/\sqrt{r^2 + \epsilon^2}$ and $\Phi_s = (3x^2 + z^2)/2$.
The energy variable can then be redefined to include the gravitational potential 
energy of the planet $E\rightarrow p/(\gamma-1) + \rho v^2 /2 + \rho \Phi_p$. 
With this definition, the energy appears more in keeping with that of standard 1D models (but with also the kinetic energy included).
Using the fact that 
$\partial \Phi/\partial t = 0$, the full energy equation (\ref{eq:energy}) can be rewritten as
\begin{equation}
\dot{E} = \dot{E}_{\rm adv} + \dot{E}_{\rm rad} + \dot{E}_{\rm work}
\end{equation}
with the three component terms described as follows.
\begin{equation}
\dot{E}_{\rm adv} = -\nabla\cdot (Ev+pv)
\end{equation}
is an advective energy term that tends to be positive and act as an energy source, heating the envelope 
and slowing accretion. Being purely the divergence of an energy flux, the advective energy is simply 
the portion of the total energy that obeys energy conservation simply by considering the continuity of 
energy moving into or out of a given volume.
This advective transport can be due to both accretion with accreted material 
adding its own energy to the envelope or recycling removing low entropy 
material and replacing it with high entropy disk gas. In the \texttt{-noshear} simulations where recycling 
is minimal, this term ought to be dominated by the accretion component, giving also some estimate of the 
relative importance of accretion vs.\@ recycling in the fiducial runs.
\begin{equation}
\dot{E}_{\rm rad} = \rho\kappa\beta(J-S)
\end{equation}
is an energy sink term due to radiative cooling given by our solution of the time-independent transfer equation. 
\begin{equation}
\dot{E}_{\rm work} = -\rho v\cdot \nabla \Phi_s
\end{equation}
is an energy term acquired by doing work against the vertical stellar and the background shear components of the potential. 
Because these components of the potential scale as $z^2$ and $x^2$ respectively, this term tends to be most relevant at large $\sim H_0$ scales.

\begin{figure*}
\plotone{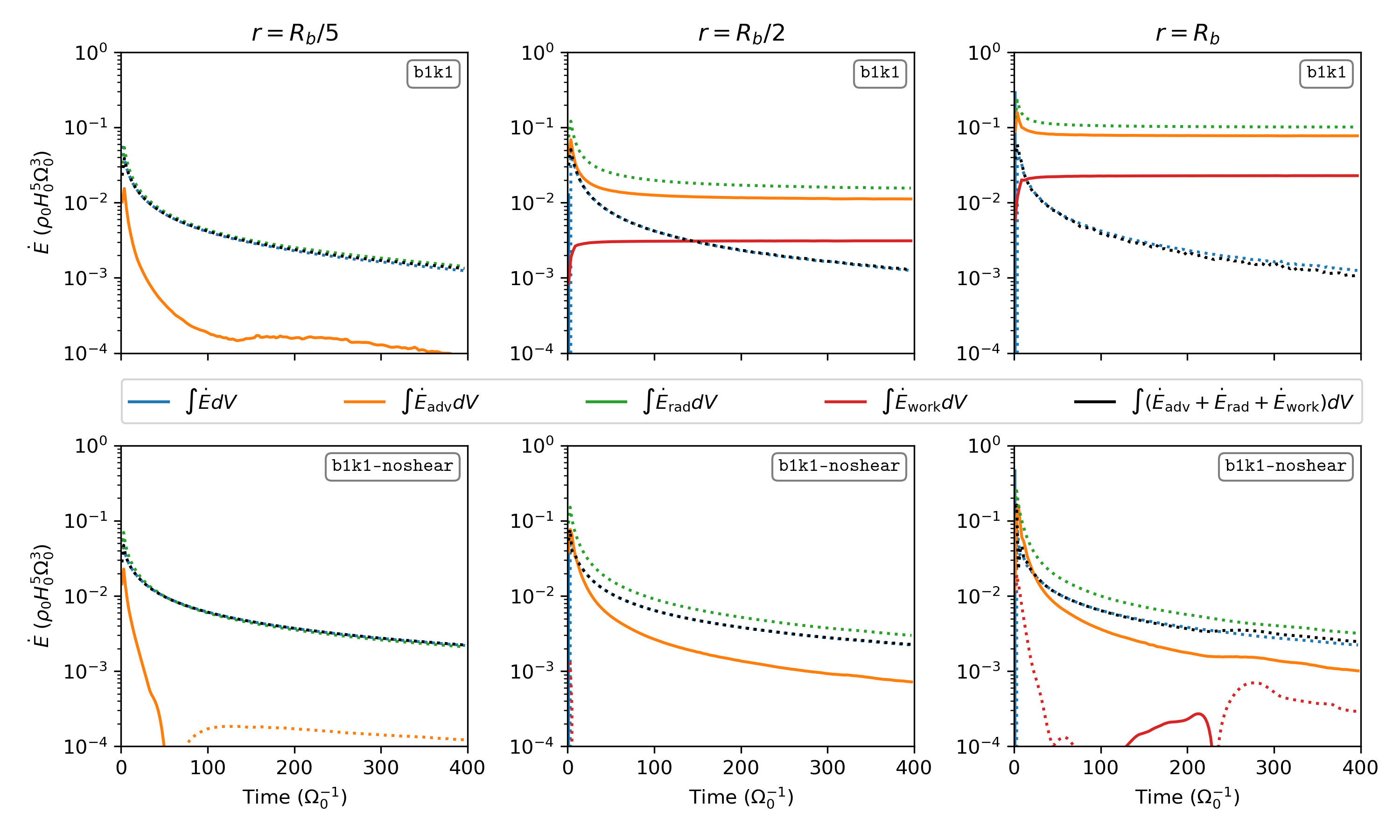}
\caption{Energy sources and sinks integrated over cubes of half-width $R_b/5$ (left), $R_b/2$ (center), and $R_b$ (right). Dotted lines are negative in sign. Top panels are taken from the fiducial \texttt{k1b1} model and bottom panels from the equivalent \texttt{-noshear} model. \label{fig:k1ebudget}}
\end{figure*}

\begin{figure*}
\plotone{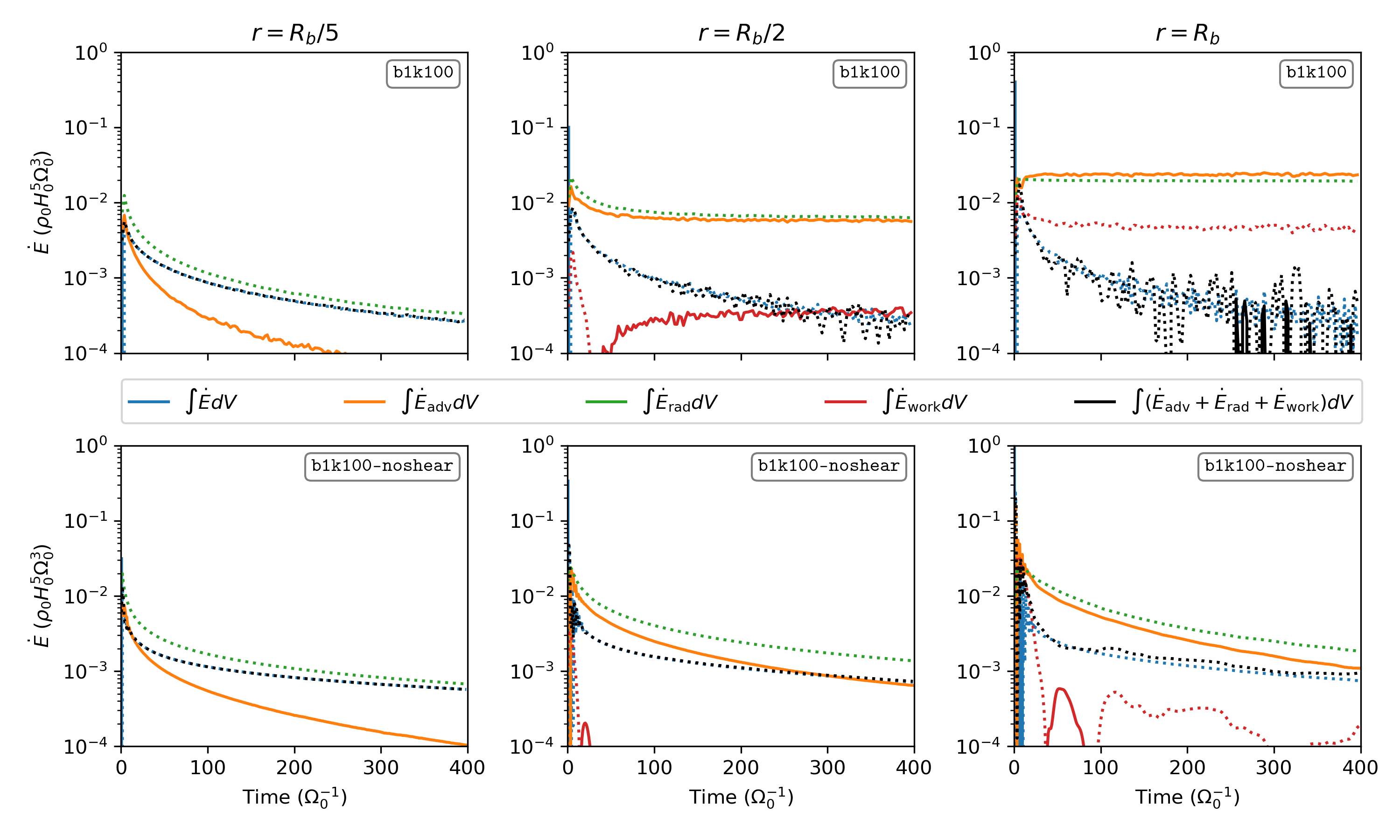}
\caption{Same as Figure \ref{fig:k1ebudget} but for the \texttt{b1k100} models.\label{fig:k100ebudget}}
\end{figure*}

In the previous section, and in keeping with the finite volume methods of simulation, we studied 
integrated mass measurements in concentric volumes centered on the planet.
Similarly here we consider the energy budget of these planet envelopes by integrating each of the three energy 
terms $\dot{E}_{\rm adv}$, $\dot{E}_{\rm rad}$, $\dot{E}_{\rm work}$ over equivalent control volumes. 
In Figures \ref{fig:k1ebudget} \& \ref{fig:k100ebudget} we plot these energy terms after integrating over control 
volumes of half-width $R_b/5$, $R_b/2$ and $R_b$. Figure \ref{fig:k1ebudget} shows the energy budgets for the 
\texttt{b1k1} and \texttt{b1k1-noshear} models while Figure \ref{fig:k100ebudget} contains the results for the optically thicker \texttt{k100}
equivalents. As a check, we compare the sum of the components  $\dot{E}_{\rm adv} + \dot{E}_{\rm rad} + \dot{E}_{\rm work}$ 
calculated by post-processing the primitive variables against the true $\dot{E}$ measured during the simulation. In all cases, the 
$\dot{E}$ measured from each of these calculations agree sufficiently well.

From the rightmost panels of Figures \ref{fig:k1ebudget} \& \ref{fig:k100ebudget} we see that on a scale comparable to the 
canonical planet envelope $R_b$, the energy budget is predominantly determined from a balance of heating from the 
advective term (orange curves) and radiative cooling (green curves). The difference between the two is the subsequent net envelope cooling (blue curves) which regulates accretion. Judging by the similarity between $\dot{E}$ curves irrespective of the choice of 
$r=R_b/5$, $R_b/2$, $R_b$ is being dominated by cooling occurring at small $r$. This is consistent with the 
idea that recycling is acting to bring the outer envelope to steady state while the inner envelope continues to 
radiative cool and accrete. 
Furthermore, as one goes to smaller radius, the difference between advection and cooling is larger. 
This suggests a point deep enough in the envelope at which radiative cooling fully dominate over advection and a Kelvin-Helmholtz like contraction can be recovered.

To be fair, the leftmost panels showing the energy budget in the very interior envelope do present a 
non-zero advective term. Though the advective flux term associated with recycling tends to be smaller than the 
net $\dot{E}$ by 1-2 orders of magnitude by the end of the simulation, there remains the question whether 
this would persist over longer timescales. 
However, it appears that this $\dot{E}_{\rm adv}$ on the interior scales cannot be even be attributed to recycling if we consider the \texttt{-noshear} runs.
After all, the advective energy 
term does not solely include energy advected due to recycling but also energy advected due to accretion. In \texttt{-noshear} 
runs where we turn off the shearing terms and have a roughly spherical accretion of material, the role of recycling is 
expected to be limited and the $\dot{E}_{\rm adv}$ ought to be predominantly due to the accretion. If the inner envelope 
energy budgets are compared to the lower panels showing the equivalent \texttt{-noshear} runs, we see the magnitude 
of $\dot{E}_{\rm adv}$ in the fiducial runs is comparable to or even lower than the $\dot{E}_{\rm adv}$. This would suggest 
that the $\dot{E}_{\rm adv}$ in the fiducial runs is not actually heating due to recycling at all but rather heating due to the accretion 
of high energy gas from larger radius. 
This leads us to conclude that at least at $R_b/5$, the energetics show no sign of being effected by recycling dynamics. 
This would be consistent with the results of Section {\ref{sec:mass}}, that suggest a recycling radius at $\sim 0.4R_b$.

An additional interesting feature of this energy budget analysis is that on the larger scales, the magnitude of the 
radiative cooling term (i.e. the luminosity), grows by one to two orders of magnitude compared to the 
actual energy liberated by Kelvin-Helmholtz contraction. In a standard quasi-static picture one would expect these 
two to be equal -- any observed luminosity would be a direct result of the contraction. In these models, the 
recycling occurring in the outer envelope is continually tapping into the fresh hot reservoir of disk gas, converting that 
energy to radiation, and adding to the luminosity from KH contraction. The result is a significantly higher luminosity than 
otherwise expected from a forming protoplanet. This could potentially make embedded protoplanets more observable 
than previously thought, although separating the signal of KH contraction from that of the recycling would be another  
confounding factor in understanding the true accretion of the planet. Stated another way: were an embedded 
protoplanet observed, the true KH contraction could be significantly slower than inferred by the simple conversion 
of observed luminosity to liberated gravitational binding energy.

\section{discussion}\label{sec:disc}
\subsection{Connection with 1D Models}
Inherent to the 3D simulation of planet envelopes is the problem that these models remain prohibitively 
expensive to run for the end-to-end formation of the planet. Nevertheless, we can develop 1D models with parameterized 3D effects and simulate for the entire planet formation timescale. 
In the process, it is also possible to mitigate some of the shortcomings of the 3D models such as the assumption of a softened gravitational potential and the lack of self-gravity. 

\begin{figure*}
\plotone{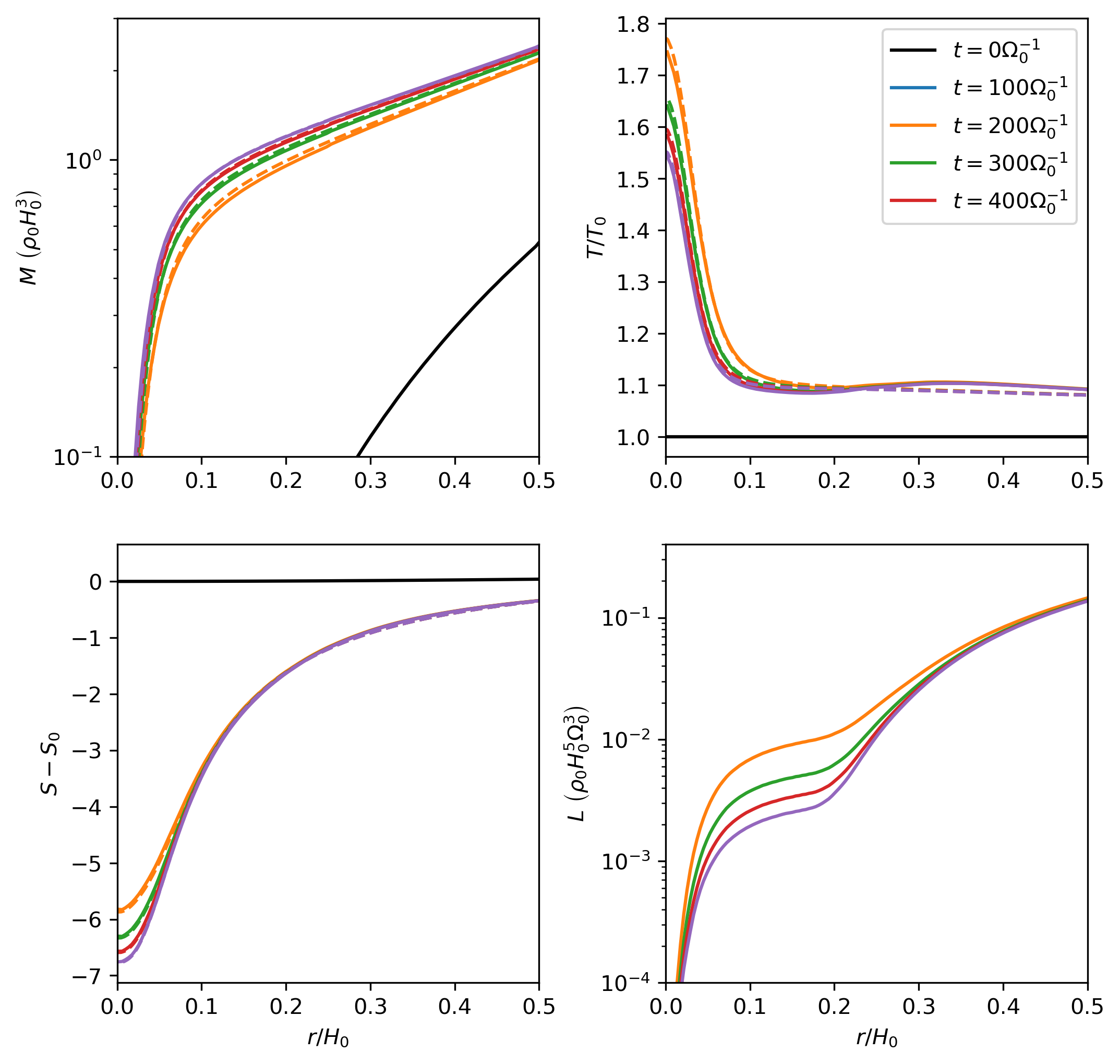}
\caption{Spherically averaged mass, temperature, entropy and luminosity profiles for the \texttt{b1k1} models (solid). Also shown (dashed) are the mass, temperature, and entropy profiles that would be obtained from applying an assumption of hydrostatic equilibrium to the measured luminosity profiles. \label{fig:1dk1}}
\end{figure*}

\begin{figure*}
\plotone{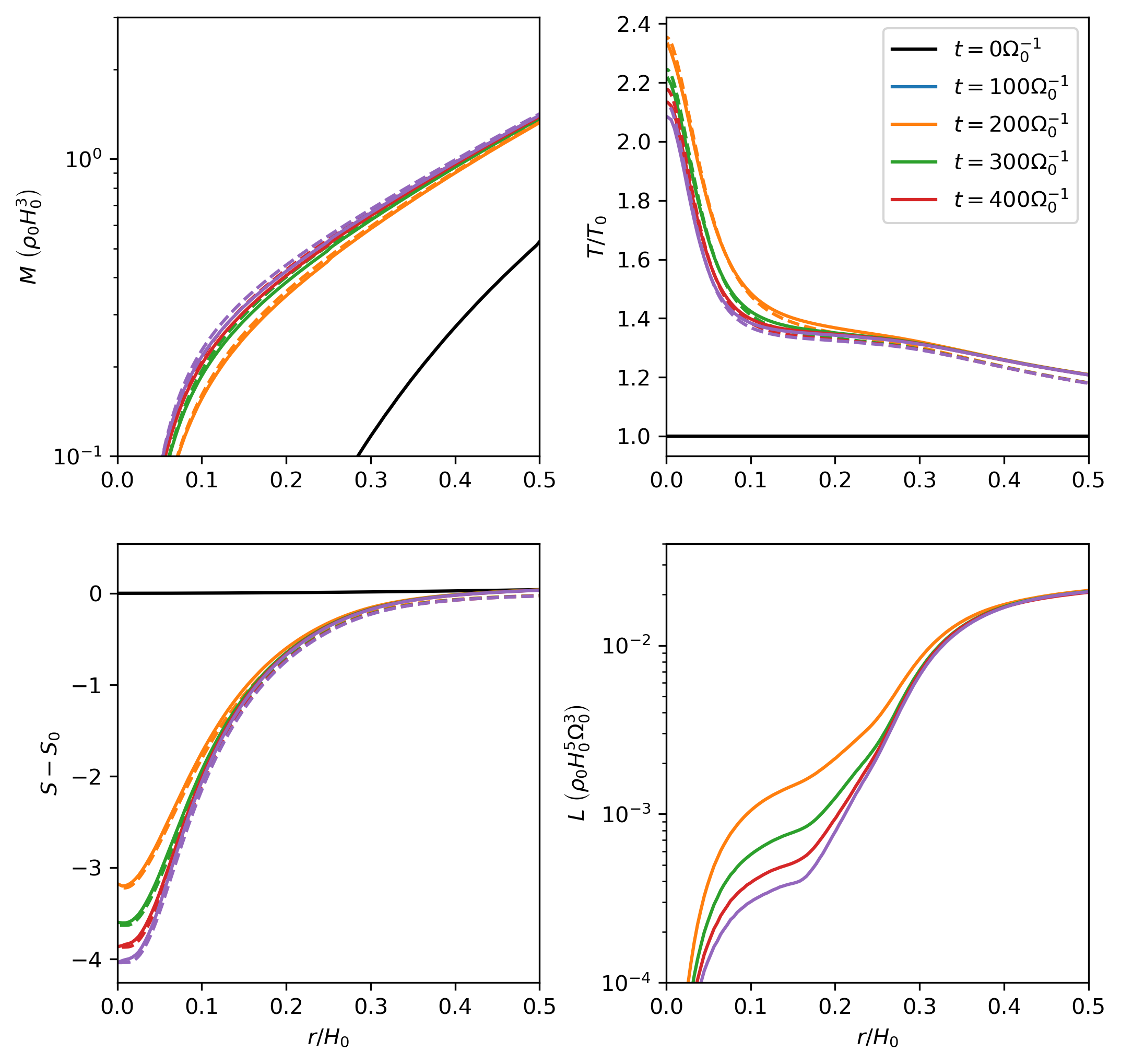}
\caption{Same as Figure \ref{fig:1dk1} but for the optically thicker \texttt{b1k100} models. \label{fig:1dk100}}
\end{figure*}

The results of the preceding sections indicate that some interior portion of the envelope models continues to cool and contract, 
suggesting that a 1D quasi-static treatment could be appropriate to model the long-term evolution. In Figures 
\ref{fig:1dk1} \& \ref{fig:1dk100} we present snapshots of the spherically averaged mass, temperature, entropy, and luminosity 
profiles derived from our 3D models. 
In the quasi-static problem, the envelope structure at a given time is entirely determined by the corresponding luminosity profile. 
This appears to also be the case in our 3D simulations and can be seen in Figures \ref{fig:1dk1} \& \ref{fig:1dk100} where also plotted are the envelope profiles retrieved by applying the hydrostatic equations to the measured luminosity profiles.
The dashed hydrostatic profiles are produced using only the spherically averaged 1D luminosity profile and 1D structure equations and match the averaged 3D results very well. 
This provides some support to the idea that a quasi-static 1D framework can be applied to adequately model the evolution.
Furthermore, if one attempts to reproduce our averaged model profiles with the standard 1D methods 
\citep{Pollack+1996, PisoYoudin2014}, the fit (not pictured) is reasonable (further suggesting that these models are participating in a sort of  
1D Kelvin-Helmholtz contraction). 
That being said, 1D quasi-static models can be improved with some minor effort, informed by the 3D envelope energetics, to better match our 3D results. In the following we describe the process and rationale 
for including these multi-dimensional hydrodynamical effects in 1D models before applying them to the long-term growth of a core under  Jupiter-like conditions as well as conditions at 0.1 AU characteristic of close-in super-Earth/mini-Neptunes.


For a first modification to 1D evolutionary models we consider the results of our mass and energy budgeting. It seems that material exterior to $\gtrsim 0.4 R_b$ is recycled efficiently, producing a net accretion rate consistent with zero in that region. 
On the other hand, material interior to this radius continues to 
cool and accrete. 
This is reflected in the bottom panels of Fig.\@ \ref{fig:vertscalar} as the location interior to which the poloidal velocity is damped to zero, in Fig. \ref{fig:mflux} as the location interior to which the envelope continues to accrete, and in the leftmost panels of Figs.\@ \ref{fig:k1ebudget} \& \ref{fig:k100ebudget} where the radiative cooling dominates the envelope energetics.
This prompts us to effectively place the canonical outer boundary of the cooling and accreting 1D envelope at $0.4R_b$ (with the scaling here being informed by the results of Fig.\@ \ref{fig:q}).

Moving the outer boundary of the canonical envelope from its standard location of $R_{\rm out} = \min\left(R_b,R_h\right)$, to $0.4R_b$ prompts an additional modification, as one must now set a value for the pressure and temperature at $0.4R_b$ to fully determine the boundary condition on the 1D model. 
One option is to say the envelope exterior to $0.4R_b$ is isentropic, reflecting the idea that the recycled outer envelope is 
continually being replenished with high-entropy disk material. 
Pressure and temperature can then be integrated along this isentrope from the background disk to derive the required values at $0.4R_b$ e.g.\@ \citet{Ali-Dib+2020}. 
This results in what we will refer to as a two-layer recycling envelope model -- an outer isentropic recycling region and an inner Kelvin-Helmholtz contracting core. 
While this is not a terrible approximation, it can be seen in both Figures \ref{fig:1dk1} \& \ref{fig:1dk100} that the entropy in the 3D models is not constant exterior to $0.4R_b$ (i.e.\@ $0.2H_0$). 
Rather, a recycled fluid element does undergo some amount of cooling in its journey from disk to planet and back. 
While mass is conserved under advection, energy is certainly not as it can be lost through radiative processes.
This is reminiscent of the results of \citet{Zhu+2021} where the thermal timescale appears to be shorter than the recycling timescale resulting in an envelope where mass is recycled exterior to $0.1R_h$ yet the thermal structure remains unaltered.
Here, the models similarly cool exterior to the fiducial $0.4R_b$ core even though mass can be rapidly replenished there.
The end result is outer steady-state envelope region that is constantly cooled and recycled in equilibrium, with the innermost 
regions being the most cooled. 
This latter point is perhaps somewhat counter-intuitive as the envelope interior is higher density and therefore has higher optical depth and slower cooling. 
However, the recycling in the inner envelope also occurs more slowly, with a given fluid element potentially making many orbits around the planet and consequently extending the time available to cool.

\begin{figure}
\plotone{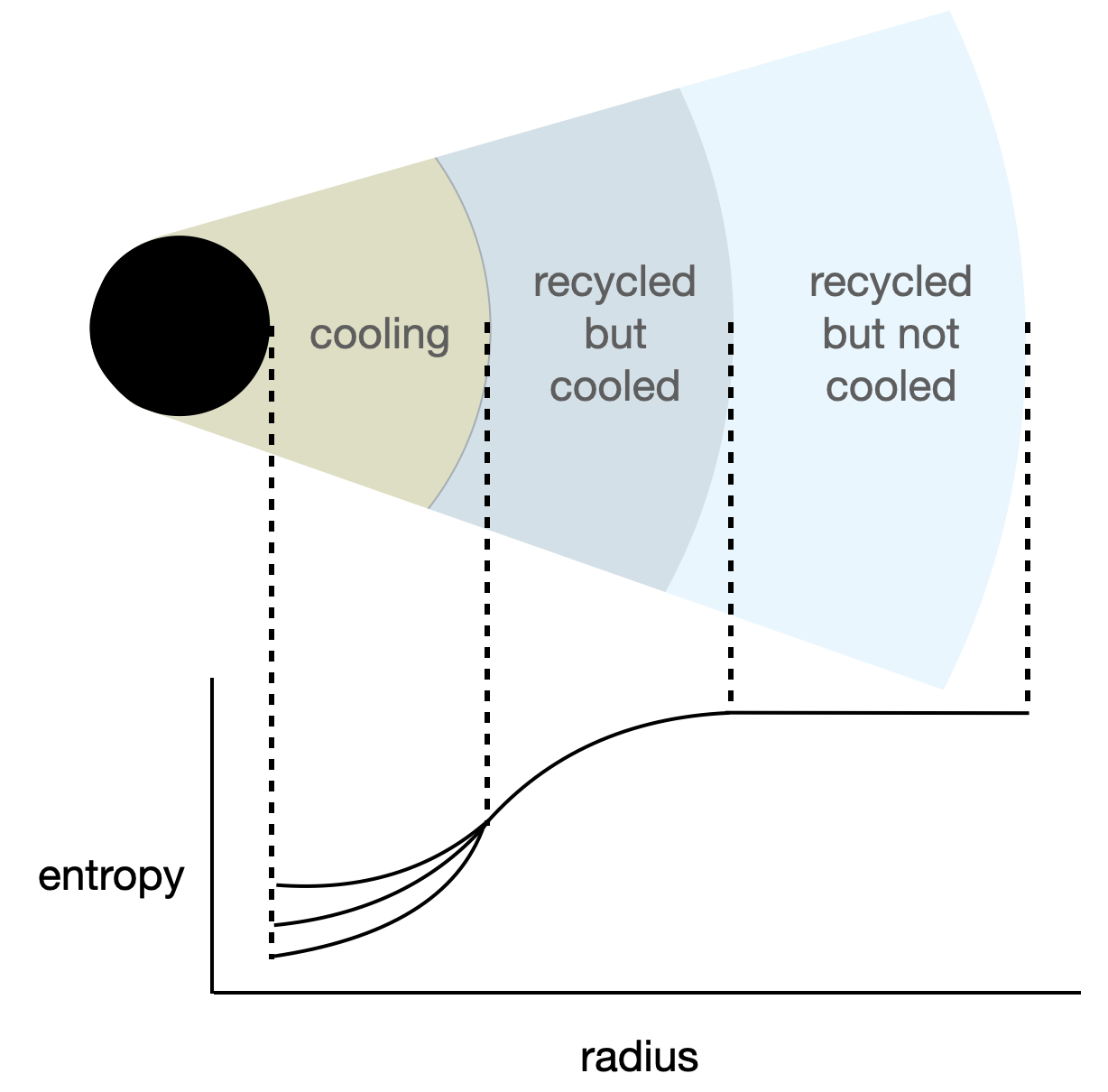}
\caption{Diagram representing the structure of our three-layer recycling 1D models.\label{fig:diagram}}
\end{figure}

While one could hope to describe this cooled but recycled region by equating the recycling and cooling times for given fluid elements, the recycling flow pattern is very much 3-dimensional and also dependent on the opacity itself making a simple analytic description for the energetics of this equilibrium challenging.
In \citet{Zhu+2021} for example, three estimates for the radial  thermal timescale are provided, all of which would give a different 1D parameterized boundary if equated to the recycling timescale here.
One effective but simple parameterization we have found to model this recycled but cooled region in 1D is to subdivide it into two regions -- an outer part where the recycling time is shorter than the cooling time, is made to be isentropic and connected to the background nebula as in the two-layer model. 
The inner part, where recycling is slower than cooling (but still steady-state), is made to evolve for a few $10^2-10^3\Omega_0^{-1}$ and taken as a fixed recycling-cooling equilibrium profile.
The resultant equilibrium profile appears fairly insensitive to the choice of the evolution time but from a physical point of view should be on the order of a characteristic recycling time. 
The only free parameter in this method is the radius at which the recycled region switches from isentropic to the cooled-recycled equilibrium. 
This parameter can be estimated from the 3D simulations and for $\kappa=1$ models is taken as $H_0$ and for $\kappa=100$ models is taken as $0.6R_b$.
Everything taken together results in a 1D three-layer recycling model as depicted in \ref{fig:diagram} which matches the 3D spherically averaged profiles better than classical quasi-static calculations and two-layer recycling models.

We recap and summarize the steps needed to reproduce our three-layer recycling models as follows.
\begin{enumerate}
    \item Integrate the nebular pressure and temperature, assuming isentropy, from the background disk\footnote{We adopt $H_0$ as nebular boundary condition radius. For $\kappa=1$ models, this means there is formally no isentropic outer layer. However, the results are insensitive to this choice and $1.5H_0$ could have been chosen for equivalent results with a truly three-layer model.} to the outer radius of the recycling-cooling equilibrium layer (estimated from 3D simulations as $H_0$ for $\kappa=1$, and $0.6R_b$ for $\kappa=100$).
    \item Evolve the quasi-static model for $\sim 10^2-10^3\Omega_0^{-1}$ to get the steady-state profile for the recycled-cooled equilibrium region.
    \item Take the pressure and temperature at $R_{\rm out} = 0.4R_b$ as an outer boundary condition to evolve the Kelvin-Helmholtz contraction of the inner envelope under the standard quasi-static framework.
\end{enumerate}

With this 3D motivated methodology, we study the long-term evolution of planetary cores at Jupiter and close-in super-Earth distances and compare to existing calculations.
To integrate the quasi-static equations for our cores we use the method of \citet{PisoYoudin2014} which integrates the usual structure equations:
\begin{equation}
\frac{dm}{dr} = 4\pi r^2 \rho
\end{equation}
\begin{equation}
\frac{dp}{dr} = -\rho \nabla \Phi_p
\end{equation}
\begin{equation}
\frac{dT}{dr} = \nabla\frac{T}{p}\frac{dp}{dr}
\end{equation}
\begin{equation}
\nabla = \min\left(\nabla_{\rm ad}, \frac{3\kappa p L}{64\pi G m\sigma T^4}\right)
\end{equation}

but makes the assumption of constant luminosity thus becoming an eigenproblem for the eigenvalue $L$ given an envelope mass. 
The sequence of time-independent structure models are then stitched together using the global energy equation
\begin{equation}
L = \frac{dE}{dt}
\end{equation}
to determine the temporal evolution.

With this methodology we explore the long-term growth of cores subject to different initial conditions and different recycling parameterizations. 
For each core we run a standard 1D model with no recycling parameterization ($R_{\rm out }= \min(R_b,R_h)$), a two-layer recycling parameterization ($R_{\rm out} = 0.4R_b$ with the boundary condition determined by an isentropic layer extending to $H_0$), and our three-layer recycling parameterization.
All 1D models remove the gravitational softening treatment and instead include self-gravity and are integrated from an inner boundary placed at the core radius assuming core density $\rho_c=3$
g/cm$^3$. 
All models also adopt $\gamma=1.4$ and $\mu=2.3$, similar to \citet{Ali-Dib+2020}.

\begin{figure}
\plotone{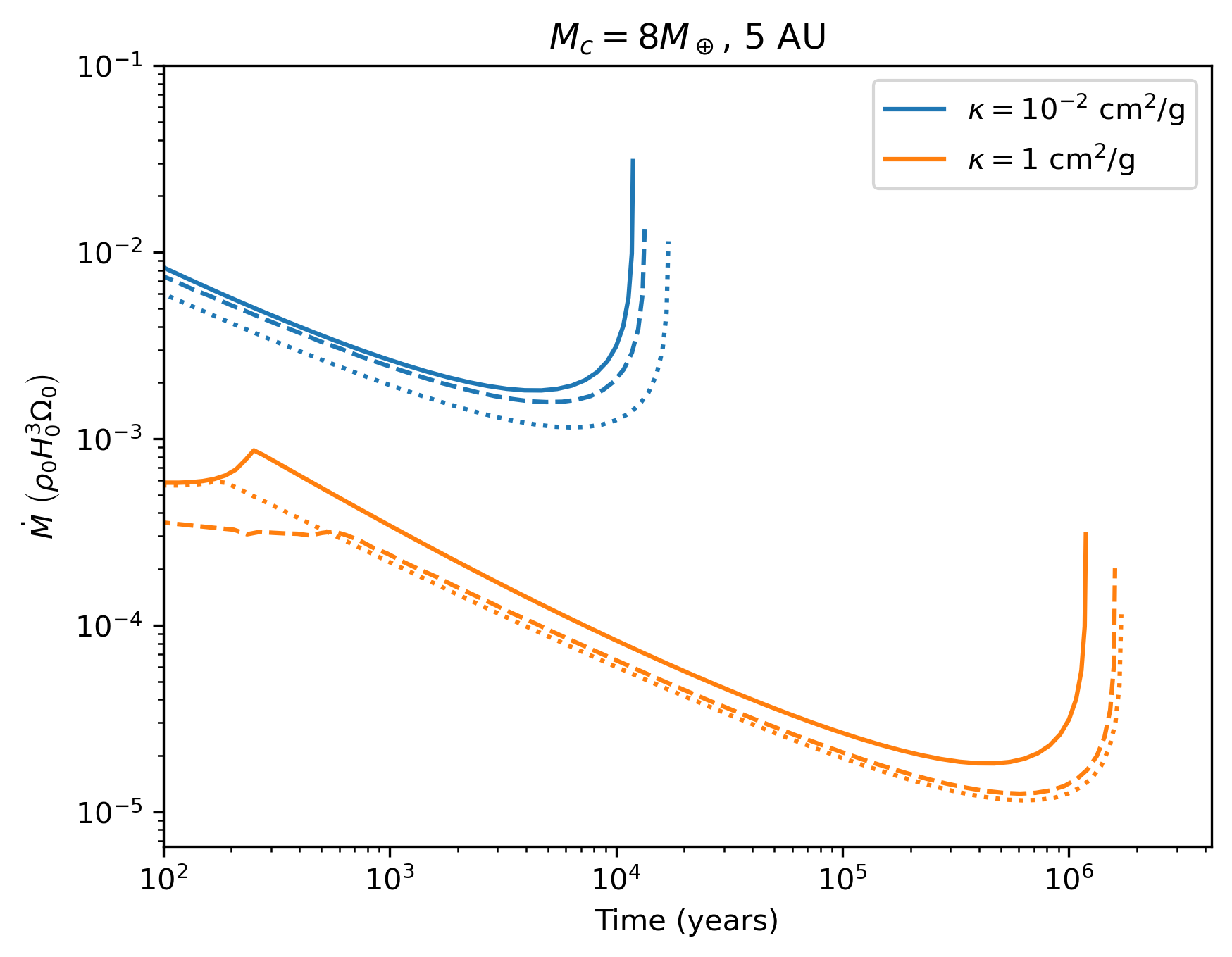}
\caption{Evolution of an $8M_\oplus$ core at 5 AU. The $\kappa_0=1$ cm$^2$/g opacity is characteristic of Jupiter-like conditions while  $\kappa_0=10^{-2}$ cm$^2$/g is only included in analogy to the 3D simulations. Solid lines are standard quasi-static models where $R_{\rm out} = \min(R_b,R_h)$, dashed lines are our three-layer recycling models, and dotted lines are an equivalent two-layer recycling model.\label{fig:jupiter}}
\end{figure}

Under Jupiter-like conditions ($M_\ast=M_\odot$, $T_0=65$ K, $\Sigma_0 = 250$ g/cm$^2$, $a=5$ AU), we display the accretion history of an $8M_\oplus$ core at $10^{-2}$ and $1$ cm$^2$/g (corresponding to dimensionless $\kappa=1$, $\kappa=100$) in Fig.\@ \ref{fig:jupiter}. 
The opacity at $1$ cm$^2$/g is characteristic of Jupiter-like conditions and reaches a phase of runaway accretion after $1$ Myr consistent with existing models. 
The low $10^{-2}$ cm$^2$/g opacity is more unphysical and shown more for the sake of completeness to apply our \texttt{b1k1} simulations.
Under these conditions, the evolution is relatively unchanged by the inclusion of a two or three-layer recycling parameterization, changing growth times by less than a factor of two. 
This is due to the fact that the envelope energetics are dominated by a high mass convective core interior to $0.4R_b$ consistent with other works \citep{Ali-Dib+2020} that employ a two-layer recycling framework and find no significant effects at $5$ AU. 

\begin{figure*}
\plotone{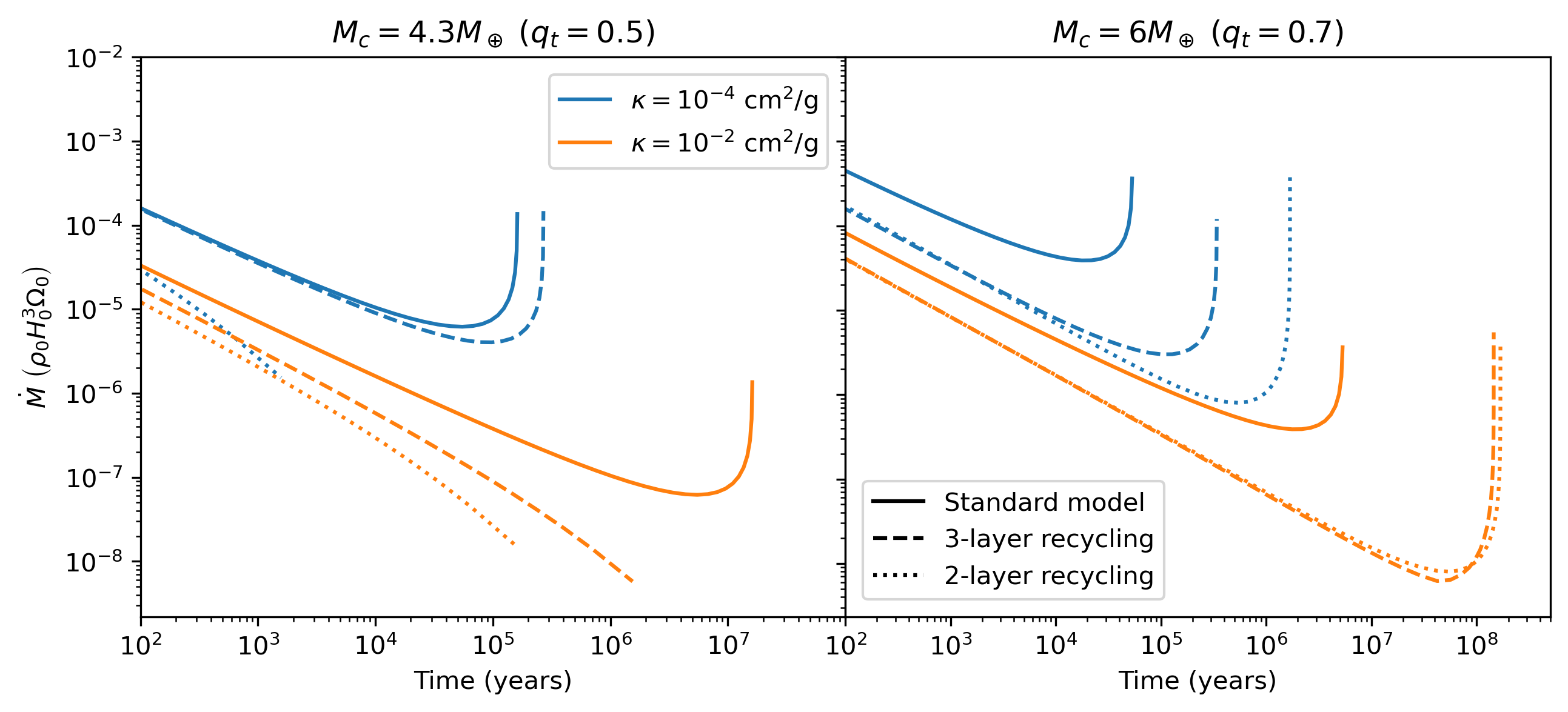}
\caption{Evolution of an $4.3M_\oplus$ core (left) and $6M_\oplus$ core (right) at 0.1 AU. Presentation is similar to that of Fig.\@ \ref{fig:jupiter}. Disk conditions at $0.1$ AU for these models are chosen as $T_0=2150$ K and $\Sigma_0 = 2.5\times 10^4$ g/cm$^2$ so that $\beta=1$. Several of the low-mass curves fail to runaway which is a consequence of the more approximate method and the imposed constant luminosity but are included for the sake of completion.\label{fig:superearth}}
\end{figure*}

In the inner disk however, recycling can act much more strongly to limit envelope growth. 
In Fig.\@ \ref{fig:superearth}, we provide the accretion history of a set of models subject to hypothetical conditions ($M_\ast=M_\odot$, $T_0=2150$ K, $\Sigma_0 = 2.5\times 10^4$ g/cm$^2$) at $0.1$ AU.
Again we adopt constant opacities corresponding to dimensionless opacities $\kappa=1,100$ which under the adopted disk parameters is $10^{-4}$ and $10^{-2}$ cm$^2$/g. 
While the $10^{-4}$ cm$^2$/g opacity might appear low relative to the assumed background temperature and density, the dust properties of these protoplanet envelopes are poorly understood and indeed models of grain settling in these environments \citep{Ormel2014,Mordasini2014} predict significantly reduced opacities.

When modeling a $4.3M_\oplus$ core under these conditions and no recycling (left panel of Fig.\@ \ref{fig:superearth}), the low opacity envelope easily runs away to gas giant within a disk lifetime. 
With a three-layer recycling model this timescale is lengthened by a factor $\sim 2$.
The other recycled models for a $4.3M_\oplus$ core reach a breaking point before the runaway phase sets in. 
This appears to be a consequence of the more approximate method and imposed constant luminosity constraint that could be addressed in the future with a coupled time-dependent treatment e.g. \citet{BodenheimerPollack1986,Pollack+1996,Ikoma+2000}.
In the meantime, we run instead a higher mass $6M_\oplus$ core which can robustly sustain a hydrostatic constant luminosity atmosphere (shown in the right panel of Fig.\@ \ref{fig:superearth}). 
The low-opacity calculation similarly runs away within a disk lifetime when no recycling is included. 
A two-layer recycling treatment however lengthens the growth time by nearly two orders of magnitude. 
At the same time, a three-layer recycling model only lengthens it by one order of magnitude. 
The appears to be a fairly generic hierarchy in that in all cases a two-layer recycling will act to lengthen the growth timescale for core and a three-layer recycling will only act to push it back to a more intermediate value.
In this recycling framework, given a fixed recycling radius, a no-recycling model gives a lower bound on the KH contraction timescale while a two-layer model produces an upper bound.

In most cases examined here, there is no significant difference between the two and three-layer treatment, but for lower opacities in the inner disk the distinction could be important like in the $10^{-4}$ cm$^2$/g, $6M_\oplus$ case. 
While these results are more pessimistic about the ubiquity of recycling halting the growth of all super-Earths than say \citet{Moldenhauer+2022}, recycling only the outer $0.6R_b$ can still slow the accretion process by a significant amount depending on the formation conditions.
In the future, a full population synthesis would do well to determine the entire scope of this process as well as the distinction between a two/three-layer treatment. Until that time, here there is at least some proof-of-concept. 

\subsection{Comparison with Other Works}
Several existing works \citep{Moldenhauer+2021, Moldenhauer+2022}  have investigated the radiation-hydrodynamics of 
protoplanetary envelopes in a manner similar to this work but obtained results different from those here. 
After several hundreds of dynamical times, the models presented in those works appear to reach a steady state 
and stop accreting. This would lead to the conclusion that super-Earth envelopes similarly be entirely recycled thus stalled 
before hitting the stage of runaway accretion. This would naturally explain the envelope mass fractions at the level of a 
few percent as a similar mass is present in the model envelopes once a steady state is reached. 
Our models do not appear to reach a steady state but rather the inner $\sim 0.4 R_b$ show quite good agreement with a 
1D quasi-static Kelvin-Helmholtz contraction. We have not been able to isolate the cause of this difference between our results 
and previous work but here examine some differences and comment on them. 

A major difference between the two works stems from the parameter space of the models studied. In our dimensionless units, the 
{\citet{Moldenhauer+2021}} models are run with $q_t=0.3$, $\kappa = 20$, and most notably $\beta=4.5\times 10^{-3}$ primarily due to a cooler protoplanetary disk model. 
This is several orders of magnitude below $\beta=1$ used here, and has a far slower cooling rate. 
In light of the results of {\citet{KurokawaTanigawa2018}}, it is entirely possible that $\beta=1$ conditions result in a quasi-static interior with $\beta \ll 1$ being fully recycled. 
Said work showed that adiabatic thermodynamics with $\beta=0$ (note this is our definition of $\beta$ and differs from the referenced work) are uninhibited by the entropy gradients and associated buoyancy barrier that arises in models with non-zero $\beta$. 
While we have tested some models with an orders of magnitude smaller $\beta$ {\citep{Bailey+2023}}, they appear essentially indistinguishable from an equivalent adiabatic simulation. 
In such a case it is difficult to determine if the cooling is truly resolved or if the radiative source term is being dominated by numerical errors. 
Hence we do not include those models here and restrict this paper only to models that are confidently converged where the energy budget analysis of Sec.\@ {\ref{sec:energy}} can be applied reliably.

Another difference between these models and others comes from the solution of the radiative transfer. While other works adopt the 
popular method of flux-limited diffusion, we calculate the formal solution to the time-independent transfer equation by 
discretizing along short characteristics \citep{Davis+2012}. Unlike M1 or diffusion approximations, 
by solving the transfer equation directly, the short characteristics method 
properly treats any anisotropies in the radiation field. 
However, in light of the results of \citet{Bailey+2023} which 
suggest only minor differences between FLD and these methods, we 
do not believe this accounts for the discrepancy.

Another potential cause lies in the choice of coordinate system and associated boundary conditions. While 
\citet{Moldenhauer+2021, Moldenhauer+2022} adopt a spherical polar system centered on the planet, our models 
utilize a Cartesian one. While we don't believe the coordinate system to have such a substantial effect on the problem, we imagine 
the associated boundary conditions could have some effect. In spherical polar one may adopt a reflecting boundary at the planetary 
surface, a good proxy for the central rocky core of the planet. In Cartesian coordinates, the models are precluded from having a 
boundary at the planet's location and must adopt some gravitational softening length instead. In having no physical boundary, 
there is an additional reservoir of material interior to the physical core radius is able to cool and accrete -- potentially lengthening 
the time to recycle. This would not entirely halt recycling however as material can still flow interior to the softening length. 
Given sufficiently long time it is possible that non-boundary models could recover the full steady state seen in spherical polar. 
If this were the case, it would actually provide a good mechanism for bifurcating the planet formation process into forming super-Earths 
at small semi-major axis and gas giants at large semi-major axis. This is because for a given core mass, the relative size difference between 
core radius and envelope radius grows as one goes further out in the disk. In this way, a core with no boundary and its reservoir of interior mass 
would be akin to a core moved further out in the disk. If the non-boundary model then showed slower recycling it would be akin to 
cores further out in the disk also exhibiting slower recycling. At Jupiter distances, perhaps this recycling would become so slow that 
runaway accretion would be recovered with the bifurcation of planet outcomes being related mainly to semi-major axis of formation. 
In the future some effort could be made to compare a diffusion 
method between coordinate systems or attempt to model a free 
reflecting sphere in Cartesian to test this.

A final hypothesis for the discrepancy in model outcomes comes from resolution effects. As seen in \citet{Bailey+2023}, 
resolution can have a significant impact on the inferred accretion rates in these types of models with under-resolution 
resulting in a false steady state. In this work, at the cost of long simulation runtime and physically small softening length, we 
have focused on obtaining converged results, running at a high enough spatial resolution to obtained converged accretion rates informed 
by convergence testing in our previous work. We emphasize that higher-order methods are essential for this kind of modeling as 
piecewise linear methods at the resolutions employed here would be insufficient to resolve the long term 
cooling we have measured.  

\subsection{Summary}
We have presented 3D models of envelope growth evaluating the role of mass and energy 
transport between envelope and protoplanetary disk for several choices of planet mass and opacity.
For the first time we have conducted an energy budget analysis to determine how robustly energy from the background disk is transported into the interior envelope through recycling.
By looking at the mass and energy budgets of the 3D models, we have determined that while there is 
obvious recycling forcing the envelope to steady state out to $\sim 0.4R_b$, this effect does not persist 
down to the inner envelope where a Kelvin-Helmholtz contraction appears to be recovered.
While the 3D models make a number of simplifying assumptions including a large softening length, 
constant opacity, etc., this allows us to ensure the measured accretion rates are converged. 
With lessons learned from these simplified 3D models, we have extended the analysis back to more realistic cases in 1D with the development of a three-layer recycling model. 
This three-layer model accounts for differences in the way mass and energy are influenced by the recycling process rather than assuming isentropy.
With these 1D models we conclude that the hydrodynamic effects would have had relatively little impact in determining the formation process of planets like Jupiter. 
For close-in super-Earths/mini-Neptunes however, recycling could have a more significant effect.
While our models suggest that only the outer $0.4R_b$ is recycled, growth times can still be delayed by one to two orders of magnitude depending on the core conditions and the recycling parameterization. 
In general, our preferred three-layer recycling models tend to delay the growth rates of protoplanet envelopes though not as much as an equivalent two-layer model.
In the future, we hope to push these 3D models further both in time and in terms of realism
to continue to test whether and under what conditions these models reach a true steady state.

\section*{Data Availability}
The data underlying this article will be shared on reasonable request to the corresponding author.

\bibliography{recycling}{}
\bibliographystyle{aasjournal}

\end{document}